\documentclass[twoside,twocolumn,9pt]{article}

\usepackage{amsmath}
\usepackage{amsfonts}
\usepackage[usenames,dvipsnames]{xcolor}

\usepackage{soul}
\usepackage{extsizes}
\usepackage[super,sort&compress,comma]{natbib} 
\usepackage[version=3]{mhchem}
\usepackage[left=1.5cm, right=1.5cm, top=1.785cm, bottom=2.0cm]{geometry}
\usepackage{balance}
\usepackage{mathptmx}
\usepackage{sectsty}
\usepackage{graphicx} 
\usepackage{lastpage}
\usepackage[format=plain,justification=justified,singlelinecheck=false,font={stretch=1.125,small,sf},labelfont=bf,labelsep=space]{caption}
\usepackage{float}
\usepackage{fancyhdr}
\usepackage{fnpos}
\usepackage[english]{babel}
\addto{\captionsenglish}{%
  
}
\usepackage{array}
\usepackage{droidsans}
\usepackage{charter}
\usepackage[T1]{fontenc}
\usepackage[usenames,dvipsnames]{xcolor}
\usepackage{setspace}
\usepackage[compact]{titlesec}
\usepackage{hyperref}

\usepackage{epstopdf}

\definecolor{cream}{RGB}{222,217,201}
\begin{document}

\pagestyle{fancy}
\thispagestyle{plain}
\fancypagestyle{plain}{
\renewcommand{\headrulewidth}{0pt}
}

\makeFNbottom
\makeatletter
\renewcommand\LARGE{\@setfontsize\LARGE{15pt}{17}}
\renewcommand\Large{\@setfontsize\Large{12pt}{14}}
\renewcommand\large{\@setfontsize\large{10pt}{12}}
\renewcommand\footnotesize{\@setfontsize\footnotesize{7pt}{10}}
\makeatother

\renewcommand{\thefootnote}{\fnsymbol{footnote}}
\renewcommand\footnoterule{\vspace*{1pt}%
\color{cream}\hrule width 3.5in height 0.4pt \color{black}\vspace*{5pt}} 
\setcounter{secnumdepth}{5}

\makeatletter 
\renewcommand\@biblabel[1]{#1}            
\renewcommand\@makefntext[1]%
{\noindent\makebox[0pt][r]{\@thefnmark\,}#1}
\makeatother 
\renewcommand{\figurename}{\small{Fig.}~}
\sectionfont{\sffamily\Large}
\subsectionfont{\normalsize}
\subsubsectionfont{\bf}
\setstretch{1.125} 
\setlength{\skip\footins}{0.8cm}
\setlength{\footnotesep}{0.25cm}
\setlength{\jot}{10pt}
\titlespacing*{\section}{0pt}{4pt}{4pt}
\titlespacing*{\subsection}{0pt}{15pt}{1pt}

\fancyfoot{}
\fancyfoot[LO,RE]{\vspace{-7.1pt}\includegraphics[height=9pt]{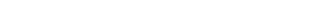}}
\fancyfoot[CO]{\vspace{-7.1pt}\hspace{13.2cm}\includegraphics{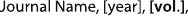}}
\fancyfoot[CE]{\vspace{-7.2pt}\hspace{-14.2cm}\includegraphics{head_foot/RF}}
\fancyfoot[RO]{\footnotesize{\sffamily{1--\pageref{LastPage} ~\textbar  \hspace{2pt}\thepage}}}
\fancyfoot[LE]{\footnotesize{\sffamily{\thepage~\textbar\hspace{3.45cm} 1--\pageref{LastPage}}}}
\fancyhead{}
\renewcommand{\headrulewidth}{0pt} 
\renewcommand{\footrulewidth}{0pt}
\setlength{\arrayrulewidth}{1pt}
\setlength{\columnsep}{6.5mm}
\setlength\bibsep{1pt}

\makeatletter 
\newlength{\figrulesep} 
\setlength{\figrulesep}{0.5\textfloatsep} 

\newcommand{\topfigrule}{\vspace*{-1pt}%
\noindent{\color{cream}\rule[-\figrulesep]{\columnwidth}{1.5pt}} }

\newcommand{\botfigrule}{\vspace*{-2pt}%
\noindent{\color{cream}\rule[\figrulesep]{\columnwidth}{1.5pt}} }

\newcommand{\dblfigrule}{\vspace*{-1pt}%
\noindent{\color{cream}\rule[-\figrulesep]{\textwidth}{1.5pt}} }

\makeatother

\twocolumn[
  \begin{@twocolumnfalse}
\vspace{1em}
\sffamily
\begin{tabular}{m{4.5cm} p{13.5cm} }

\includegraphics{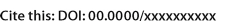} & \noindent\LARGE{\textbf{Vesicle formation induced by thermal fluctuations}} \\
\vspace{0.3cm} & \vspace{0.3cm} \\

 & \noindent\large{Andreu F. Gallen$^{1}$, J. Roberto Romero-Arias$^{2}$, Rafael A. Barrio$^{3}$, and Aurora Hernandez-Machado$^{1,4}$ }\\

\includegraphics{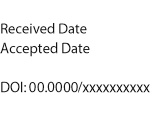} & \noindent\normalsize{ 
The process of fission and vesicle formation depends on the geometry of the  membrane that will split. For instance, a flat surface finds it difficult to form vesicles because of the lack of curved regions where to start the process.
Here we show that vesicle formation can be promoted by temperature, by using a membrane phase field model with  Gaussian curvature. We find a phase transition between fluctuating and vesiculation phases that depends on  temperature, spontaneous curvature, and the ratio between bending  and Gaussian moduli. We analysed the energy dynamical behaviour of these processes and found that the main driving ingredient is the Gaussian energy term, although the curvature energy term usually helps with the process as well. We also found that the chemical potential can be used to investigate the temperature of the system.  Finally we address how temperature changes the condition for spontaneous vesiculation for all geometries, making  it happen in a wider range of values of the Gaussian modulus.
} \\

\end{tabular}

 \end{@twocolumnfalse} \vspace{0.6cm}
 ]
  
\renewcommand*\rmdefault{bch}\normalfont\upshape
\rmfamily
\section*{}
\vspace{-1cm}

\footnotetext{\textit{$^{1}$Departament Física de la Matèria Condensada, Universitat de Barcelona, E-08028 Barcelona, Spain. }}
\footnotetext{\textit{$^{2}$Instituto de Investigaciones en Matem\'aticas Aplicadas y en Sistemas, Universidad Nacional Aut\'onoma de M\'exico, 01000 Ciudad de M\'exico, Mexico. }}
\footnotetext{\textit{$^{3}$Instituto de  F\'isica, U.N.A.M., 01000, Ap. Postal 101000, M\'exico D.F., M\'exico }}
\footnotetext{\textit{$^{4}$Institute of Nanoscience and Nanotechnology (IN2UB), 08028 Barcelona, Spain}}

\section{Introduction}
Lipid vesicles, also known as liposomes or simply vesicles, are structures that are crucial for many biological processes, as transport vesicles \cite{rothman1996protein}, secretory vesicles \cite{njus1986bioenergetics}, and endocytosis \cite{silverstein1977endocytosis}. Many of these biological systems require to create and destroy vesicles constantly, e.g. the Golgi apparatus \cite{Miserey,rothman1994}, the synaptic system \cite{miesenbock1998,Parkar}, or enveloped viruses \cite{Dharmavaram,Eckert}. Even red blood cells are known to lose membrane when they become old through shedding of microvesicles and leading to a change of their surface-volume ratio \cite{alaarg2013red}. Thus, there is great interest in gaining insight on the dynamics of vesicle formation, also referred to as vesiculation. Vesicles are also produced artificially, either from  preformed bilayers or by building them directly from an organic mixture of lipids in contact with an aqueous medium \cite{has2021vesicle}, using some electroformation \cite{angelova1986liposome} or thin-film hydration methods\cite{bangham1965diffusion}.

In a previous article \cite{rueda2021gaussian}, the vesiculation of a membrane tube was studied  by introducing the Gaussian curvature energy term. This Gaussian term is necessary to study topological transitions on membranes and has usually been neglected \cite{siegel2008gaussian}. The membrane tube geometry helped the fission process and leads to the membrane splitting into multiple vesicles \cite{rueda2021gaussian}. However, there are some membrane geometries that instead of helping vesicle formation process they hinder it like, for example, a perfectly flat membrane.

A perfectly flat membrane is a geometry that hinders this budding process as it lacks any curved region. Without any point more curved than the other, even if budding is energetically favourable there is no preferred point from where to start the membrane budding process. This, in turn, stops vesiculation, even in the case where generating vesicles is energetically favourable. However, in nature vesiculation happens in wide variety of situations and with plenty of membrane geometries for the initial membrane. It is known that vesicle formation is a temperature-dependent process \cite{silverstein1977endocytosis,scott1979plasma} and that temperature gradients affect the phase behaviour  heterogeneous membranes \cite{atia2014theoretical} and can change the lipid phase between fluid and gel states \cite{dimova2014recent}. In experimental works, it has been shown that the size of the vesicles depend on preparation methods, the type of lipid used and the temperature \cite{has2020comprehensive} and  even  the temperature has been using to produce fission of vesicles in Giant Unilamelar Vesicules (GUVs)\cite{leirer2009phase} considering  the first order phase transition between the gel and fluid states in the lipids.
{\color{black} In live organisms we also have temperature dependence many processes like synaptic signalling, endocytosis, and in vesicle trafficking \cite{chanaday2018time}.}
 
The processes beyond membrane fission has been challenging. There are some works in microscopic modelling with molecular dynamic simulations\cite{laradji2004dynamics,yang2009computer,yang2012computer}, but in the mesoscopic modelling scales little has been done, and there is a bit  understanding on topological transitions \cite{roadmapBassereau, CampeloReview}. In mesoscopic models, one needs to consider the energy of the membrane as a continuous surface, instead of going molecule by molecule. Thus, in this work,  we study some mechanisms that promote --or even induce-- the formation of a vesicle from a non-favourable geometry.

To promote the vesiculation from a completely flat membrane, one introduces the effect of finite temperatures to the previously mentioned model\cite{rueda2021gaussian}.
We develop a three dimensional model to study the dynamics of spontaneous vesicle formation from a flat membrane by adding a thermal noise. 
{\color{black} This model will be accurate for vesicles or regions of a membrane where there are no proteins present.}
With these considerations, one shall expect to deform the membrane enough and start the vesiculation process. However, it is not as simple as that, as there is an energy barrier that is impeding the flat membrane to change shape and form buds. Thus, we assemble a phase diagram between fluctuation and vesiculation states and we find that the topological stability of the membrane depends on the bending and Gaussian moduli, the spontaneous curvature and the temperature of the system. Also, we find that temperature is only needed to trigger the process but that when one has a curved enough region the vesiculation process will go on, even when turning the temperature back to zero.
Surprisingly, we see no topological transitions for positive Gaussian modulus.
One would expect the membrane to form tunnels and channels but the topology of the system remains unchanged even for excessively large temperatures.\\

\section{Model}

The model follows the Canham-Helfrich free energy and its phase-field approaches used for fission of membrane tubes\cite{rueda2021gaussian}. The  Canham-Helfrich free energy is written as 
\begin{equation}
F=\int_A \Big( \frac{\bar\kappa}{2}(C-c_0)^2+\bar\kappa_GK+\gamma_A\Big)dA \, + \int_V \Delta p \, dV\text{,}
\label{eq:HelfrichHam6}
\end{equation}
where $\bar\kappa$ and $\bar\kappa_G$ are the bending  and Gaussian moduli, respectively. $C$ is the total curvature,  $c_0$ is the spontaneous  curvature and $K$ is the Gaussian curvature. The parameters $\gamma_A$ and $\Delta p$ are Lagrange multipliers that ensure the conservation of the area and the volume respectively. These last can be understood as an effective surface tension and an osmotic pressure that works against changes of area and volume, respectively. 

Usually, by using the Gauss-Bonnet theorem one can avoid the Gaussian curvature contribution to the energy, but in this  work, we consider the Gaussian contribution since we are interested in studying the formation of vesicles. The Gauss-Bonnet theorem states that the integral of the Gaussian curvature over a surface is proportional to the surface Euler characteristic $\chi(\Omega)$, which is a topological invariant \citep{do2016differential}. The Euler characteristic describes the system by the number of holes $g$ and objects $N$ like  $\chi(\Omega)= 2 (N-g)$. In general, one can describe a system that increases the number of objects and generate vesicles with $\bar \kappa_G <0$  and a system  that increases the number of holes when  $\bar \kappa_G >0$\cite{roadmapBassereau}.

From the free energy of equation (\ref{eq:HelfrichHam6}), one can compute the minimum energy necessary to form a spherical vesicle from a planar membrane.
The energy required, if we neglect the bilayer thickness effect\cite{has2021vesicle}, reads 
\begin{equation}
    f = (2\bar\kappa+\bar\kappa_G)\frac{A}{R^2},
    \label{eq:vesicle_energy}
\end{equation}
were $R$ is the radius of the vesicle, $A$ is its area. To have vesicles generated spontaneously, we need this energy contribution to be negative.
Thus, one can obtain that for the vesicle formation to be spontaneous, one requires $- \bar\kappa_G >  2\bar\kappa$.
Therefore, one should find spontaneous vesiculation depending on the values of $\bar\kappa_G$ and $\bar\kappa$. 
The values of the bending modulus are estimated from elastic properties of the membranes \cite{roadmapBassereau}. 
However, the Gaussian modulus needs an indirect technique for measurement. An experimental\cite{siegel04} and theoretical\cite{Hu} value of Gaussian modulus is estimated as $\bar \kappa_G \approx -15 K_BT$. In a previous work\cite{rueda2021gaussian}, we studied a phase diagram on vesicle formation and we found three possible stable states that corresponds to: i) membrane fission when the negative Gaussian modulus $\bar\kappa_G$ is much larger than the bending modulus $\bar \kappa$; ii) no fission membrane when $\bar \kappa_G$ is negative but not  much larger than $\bar\kappa$; and iii) multiple self-connected membrane for positive $\bar \kappa_G$ in which many holes are developed in a single continuous membrane. 

In the phase diagram $(\bar\kappa_G,\bar\kappa)$, one finds that for negative Gaussian modulus there is a transition between spontaneous vesiculation state and a topologically stable state defined by the line $\bar\kappa_G = - 2\bar\kappa$. As one can compute the energy cost to generate a vesicle, one may compute the energy cost necessary when one has a Gaussian modulus that does not comply to that condition. Thus, in order to generate a spherical vesicle of  area $A=4\pi R^2$ with radius $R$ and $\bar\kappa_G <0$, the energy required is $ f = 4 \pi (2\bar \kappa - | \bar \kappa_G | ) $. 
{\color{black} This energy $\Delta E$ is the minimum energy required for generating a vesicle.}
Introducing now a spontaneous curvature $c_0$ \cite{jung2001origins}, we get that the energy barrier is $ \Delta E = 4\pi (2\bar\kappa+\bar\kappa_G)(1-R \, c_0)^2$.
{ Fission is a thermally activated process with a characteristic time-scale of around $\tau= t_0 e^{\Delta E /K_BT}$ where $t_0$ is the typical time scale of a thermal fluctuation that depending on the system  goes from $1 ns$ in membrane tubes to $1 ms$ on the Golgi apparatus\cite{campelo2017sphingomyelin,morlot2012membrane}.

Thus the system will follow an Arrhenius law. 
For long enough times, vesiculation will occur in any case that has a thermal energy  $K_BT$ larger than $\Delta E$. 
If $T$ is too small the time required for vesiculation will be much longer than the simulation time. }
Therefore, depending on whether the thermal energy, $E_T = k_BT$, is bigger or smaller than this energy barrier, one obtains spontaneous vesicle formation or not.
This is what will distinguish between the two phases that we study in this paper, the fluctuating and the vesiculating one. 

Vesiculation is an irreversible process, as it increases the entropy of the system and  we can ensure that for a time long enough the system will generate vesicles. Therefore,  for a system with a temperature  $T$ the condition for spontaneous vesicle formation will be
\begin{equation}
    - \bar\kappa_G > 2\bar\kappa - k_BT.
     \label{eq:kg_vs_k_T}
\end{equation}
With this formulation,  the membrane may exhibit spontaneous fission for new values of $\bar\kappa_G$ that could not vesiculate before.\\

{\bf Membrane phase field model}

{\color{black} Phase field models are used in problems where tracking the position of an interface can be difficult or computationally costly. These models have been used to study membranes successfully in the past for both equilibrium and dynamic systems \cite{campelo08,Misbah,LazaroSM1,Du}.}
One can write the free energy of the membrane from equation (\ref{eq:HelfrichHam6}) as a function of an order parameter $\phi$ and its derivatives \cite{campelo06}.
This order parameter $\phi$ will define if a given point in space corresponds to:  the membrane, the inner, or the outer fluid of the vesicle. 
We can obtain the dynamic equation by performing the functional derivative of both contributions to the free energy and write the temporal evolution. 
{ The dynamic equation is a Ginzburg-Landau formalism (model B \cite{hohenberg1977theory}) for a conserved order parameter as }
\begin{equation}
\frac{\partial \phi}{\partial t} = \nabla^2 \left( \frac{\delta {F_{C_0}} }{\delta \phi} + \frac{\delta {F_K} }{\delta \phi} \right),
\label{eq:dynamic_equation1}
\end{equation}
in which the mean and spontaneous curvature $F_{C_0}$ contribution  reads like \cite{campelo08}
\begin{equation}
  {F_{C_0}} = \kappa \int_\Omega  \Big( (\phi-\epsilon C_0)(\phi^2-1)- \epsilon^2 \nabla^2 \phi  \Big)^2 dV,
  \label{eq:FC0}
\end{equation}
where {\color{black} $\epsilon$ is the width of the interface,} $\kappa = 3\bar{\kappa}/4\sqrt{2}\epsilon^3$ and $C_0 =c_0/\sqrt{2}$. 
Similarly, the Gaussian contribution is\cite{rueda2021gaussian}
\begin{equation}
      {F_K} = \int \bar \kappa_G K dS = \bar \kappa_G \frac{3}{4\sqrt{2}\epsilon} \int \sum_{i<j} \frac{(1-\phi^2)^2}{2} \Big(Q_{ii} Q_{jj}-Q_{ij}^2\Big)dV,
  \label{eq:FG}
\end{equation}
where Gaussian modulus $\kappa_G=3 \bar \kappa_G/(4\sqrt{2}\epsilon)$ and $Q_{ij}$ is  curvature tensor (see Supporting Information for the exact expression). The dynamical equation (\ref{eq:dynamic_equation1}) can be written explicitly in terms of the phase field order parameter $\phi$ as 
\begin{equation}
\begin{split}
    \frac{\partial \phi}{\partial t} &=  \kappa\nabla^2 \Big( (3\phi^2-1-2\phi\epsilon C_0)\Phi-\epsilon^2\nabla^2\Phi + \gamma_A\nabla^2\phi \Big)\\
    &-\kappa_G \nabla^2 \left( \frac{12\phi}{1-\phi^2} T_1 + \frac{2(3\phi^2+1)}{(1-\phi^2)^2} T_2   \right),
\end{split}
\label{eq:dynamic_equation}
\end{equation}
where $ \Phi = (\phi-\epsilon C_0)(\phi^2-1)- \epsilon^2 \nabla^2 \phi $. 
The terms $T_1$ and $T_2$, which can be found in the Supporting Information, are computed from various derivatives of $\phi$ \cite{rueda2021gaussian,campeloTh}.

{\color{black} Even though in eq. (\ref{eq:dynamic_equation}) we still keep the Lagrange multiplier for area $\gamma_A$ in this membrane geometry of an infinite flat surface we do not need to impose the conservation of surface area. 
The membrane will behave as if connected to an area reservoir and thus we will not introduce $\gamma_A$ in this numerical implementation.
Regarding the volume, local volume conservation is always ensured thanks to the diffusive behaviour of eq (\ref{eq:dynamic_equation}) as we have all the dynamics inside the Laplacian.} \\

{\bf Numerical Implementation}

The equation (\ref{eq:dynamic_equation}) is highly non-linear with the order parameter and thereforeshall solve it numerically. 
It has been already corroborated that a  Euler method {(forward Euler)} is reliable to solve this system provided the time step $\Delta t$ is small enough \cite{rueda2021gaussian}. 
We work with a lattice of distance $\Delta x$ between the points, which is constant along the system and equal for each dimension $(x,y,z)$. 
{ 
The scales chosen for the implementation were $\Delta x = 1$ for space and $\Delta t = 3 \cdot 10^{-4}$ for time. 
}

Thus, we define the initial conditions of the  order parameter $\phi(x,y,z)$ and compute the temporal evolution of the dynamic eq. (\ref{eq:dynamic_equation}). 
We should point out that there is complication when  computing the contribution
\begin{equation}
    \frac{12\phi}{1-\phi^2} T_1 + \frac{2(3\phi^2+1)}{(1-\phi^2)^2} T_2 ,
    \label{eq:complexphi}
\end{equation}
as in the bulk $(1-\phi^2)\to 0$ and this would diverge. We avoid this by adding a very small  imaginary contribution to $\phi$ and implementing the integration using the residue theorem over the upper complex plane for equation (\ref{eq:complexphi}).

In the dynamical equation (\ref{eq:dynamic_equation}), the bending and the Gaussian contributions can be differentiated by their energy moduli $\kappa$ and $\kappa_G$, respectively. Thus, the Lagrange multiplier for the volume is not necessary since the dynamic equation is written conservatively following the diffusion equation\cite{campeloTh} (also known as model B). On the other hand,  the Lagrange multiplier for the area conservation $\gamma_A$, that is necessary when working with finite closed membranes,  will not be computed and considered as a constant, because we work with a reservoir for the surface area and the volume.
\\

{\bf Thermal fluctuations}

The thermal agitation of the membrane is modelled by white noise. 
This white noise represents the displacement generated by the random impacts of molecules on the membrane due to their random Brownian motion. 
Those impacts displace constantly the membrane from its average position moving different points of the membrane to different directions with various displacements. 
Therefore, we consider that thermal fluctuations are delta-correlated thorough space and time. 
This means that when one compares two different points of space, the fluctuations over a long enough time require to have no correlation between them, therefore being independent. 
It is necessary to have a noise that its intensity independent of time and space.
From the Fluctuation Dissipation theorem, we can obtain the relation between the noise variance and the temperature via its auto-correlation which, for a noise $\xi$,  takes the form
\begin{equation}
\langle \xi(\mathbf{x},t) \xi (\mathbf{x}',t')\rangle = \frac{2k_BT}{\bar\kappa}\delta(\mathbf{x}-\mathbf{x}')\delta(t-t'),
\label{eq:autocorr_xi}
\end{equation}
where $k_B$ is the Boltzmann constant and $T$ the temperature.
Its inverse dependence on the bending modulus $\kappa$ comes from the fact that a higher rigidity reduces the thermal fluctuations of a membrane. 
his noise will be acting on the membrane position through the order parameter $\phi$. 
The white noise has a mean $\langle \xi \rangle=0$, a standard deviation $\sigma$ and Gaussian probability distribution. 
The variance is the second moment of a single random variable ${\xi}$ is defined by
$    \sigma^2 \equiv \langle [ \xi - \langle \xi \rangle]^2\rangle $
and is the measure of how much the values of $\xi$ deviate from the mean value $\langle \xi \rangle$ \cite{gardiner1985handbook}.
The intensity of the noise can be measured with the variance $\sigma^2$ {and from the Fluctuation Dissipation theorem, we know that} for a thermal noise this variance is  proportional to the temperature by
\begin{equation}
   \sigma ^2= \frac{2k_BT}{\bar\kappa}. 
\end{equation}

We can identify in equation (\ref{eq:autocorr_xi}) that the magnitude of the auto-correlation is the variance $\sigma^2$ which can be modulated with a  Gaussian distribution.

Now, to add this delta-correlated noise into the system, we consider that the membrane is obtained for the points in space defined by $\phi=0$ and the interface relaxes fast to the hyperbolic tangent profile. Thereby, one can simply add a noise to the $\phi$ value around the points of the diffuse interface where $-1<\phi<1$.
This would be written like 

\begin{equation}
\begin{aligned}
\frac{\partial \phi(\mathbf{x},t)}{\partial t}=   \nabla^2 \Big(\frac{\delta F_{C_0}}{\delta \phi} + \frac{\delta F_K }{\delta \phi} \Big)
+ \xi(\mathbf{x},t)(1-\phi^2)/\epsilon^2,
\end{aligned}  
\end{equation} 
where the noise disturbs only the membrane while most of the bulk remains undisturbed.

Nonetheless, the noise has a Gaussian distribution with a mean zero, the conservation of volume and area can be violated. 
Even if the mean of the noise is zero for short time-scales we can find that the overall contribution of the noise is non-zero. 
Thus, one has to be careful when introducing temperature to a membrane system to ensure that the area and volume are conserved. 
Thus, to facilitate fission, we study a flat membrane with a reservoir of surface area and volume. 
It is known that individual closed vesicles should not be able to change area and volume \cite{helfrich73}. 
This is more difficult to ensure for individual vesicles than it would be for the system as a whole. 
The vesicles size depend on the spontaneous curvature of the membrane \cite{seifert1991shape} in most cases the resulting vesicles are small \cite{alaarg2013red} and thus the membrane fluctuations cannot be taken as small for their length-scale. 
With their sizes the area and volume of the vesicles slightly change and fluctuate over time. 
This could be addressed by adding complex Lagrange multipliers to the model, however with the model as is one can study whether vesiculation happens or not, its energy, and the dynamics that bring a membrane to vesiculate. 
The proper modelling of the vesicles after they have scissioned from the mother membrane would add little or nothing given our goals.

{ The relaxation rate, and thus the spectrum, of membrane phase field models has already been studied on \cite{Lazaro2017}. 
There the authors found that the power spectrum falls with the wavector $q$ as $\langle | h_q|^2\rangle \propto q^{-5}$ where $h_q$ is the deformation of the membrane in the Fourier space.
The expected  $\langle | h_q|^2\rangle \propto q^{-4}$ is recovered when coupling the membrane to the Navier-Stokes equations \cite{Lazaro2017}.
In the present model we do not introduce the Navier-Stokes equation.
Here we want to study whether vesiculation happens or not, and under which conditions of energy modulus and temperature. 
The presented model thus should not be used to study the actual dynamics in real time without the addition of the hydrodynamic contribution.
{ 
Our model in eq (\ref{eq:dynamic_equation1}) is taking into account that the system is fluid and that the relaxation from a perturbation requires a conserved order parameter \cite{hohenberg1977theory} giving a diffusion equation.
It is known that in the case of phase transition dynamcics when you use a diffusive dynamic equation you get an extra $q^{-1}$ contribution on the power spectrum \cite{gunton1983introduction,jasnow1987crossover}.
The introduction of hydrodynamics cancels this contribution giving a $\langle | h_q|^2\rangle \propto q^{-4}$ due to the Oseen tensor \cite{Lazaro2017}.

This does not invalidate any of the results we show. The drawback of this method is that we cannot study space-correlations without the complete model (including hydrodynamics). 
Moreover,} we expect that the behaviour of the membrane when uncoupled with the fluid will be accurate for the low-$q$ regime, which is the regime of big deformations and as such of vesiculation. 
Our current model will have less fluctuations in the high-$q$ regime, and we expect the membrane to have less ripples and small deformations than an actual membrane.

}

\section{Results}
{\bf Negative Gaussian modulus}

We focus on simulating flat infinite membranes where, as discussed earlier, there is an energy barrier that needs to be overcomed  to reach fission. With the white noise on the membrane position, we can mimic the effects of temperature. 
Thus, with enough thermal energy, we expect the membrane to produce fission. However, this depends on the interplay between temperature and the bending and Gaussian rigidities $\kappa$ and $\kappa_G$, respectively. 
That means that, if the temperature is not large enough, for a given  $\kappa$ and $\kappa_G$, one finds that even if the membrane can fluctuate, the fluctuations will not be enough to bend the membrane enough to produce a bud.

\begin{figure}[ht!]
\begin{center}
\includegraphics[width=.49\columnwidth]{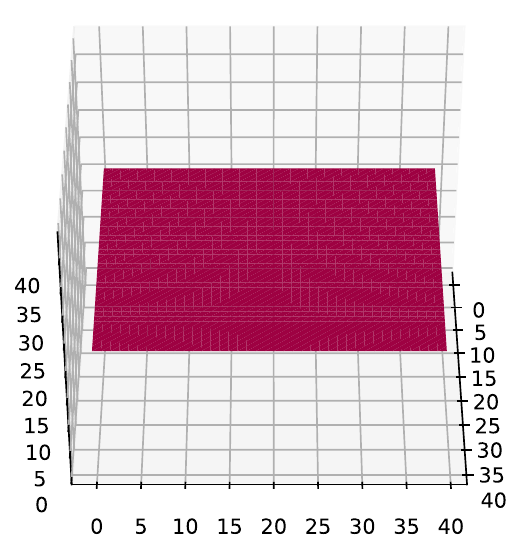}
\includegraphics[width=.49\columnwidth]{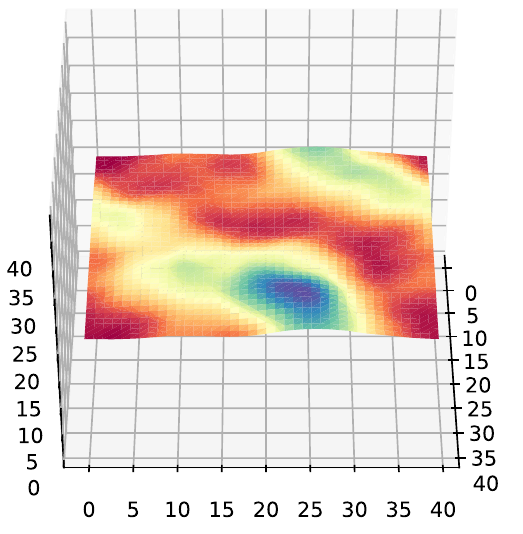}
\caption{\textbf{Left:} Snapshot of a flat membrane under conditions where fission is energetically favourable.  However due to the starting geometry, we see no change over time on the membrane shape. The parameters used are $T=0$, $\epsilon=1$, $\kappa=1$, $\kappa_G=-10$ and $C_0=-0.5$. \textbf{Right:} Snapshot of a fluctuating flat membrane under conditions where fission is energetically favourable and with low temperature.
The parameters used are  $T=1.33\cdot10^{-5}$ in internal units (IU) defined in the text, $\epsilon=1$, $\kappa=1$, $\kappa_G=-2$ and $C_0=-0.25$. The color map of the membrane represent its local height. Red color means the membrane goes down and blue color means the membrane goes up.
}
\label{FIG:flat_membrane_t0}
\end{center}
\end{figure}

In the left side of Figure \ref{FIG:flat_membrane_t0}, we show a simulation with no noise at temperature $T=0$. {}
The membrane remains flat and no deformation occurs no matter how long the simulation runs. When one adds temperature to the flat membrane, one could start to see fluctuations that take the shape of something like the right side of {Figure \ref{FIG:flat_membrane_t0}}. In this case, the thermal fluctuations are not big enough to produce vesicle formation and one can define this as the fluctuating regime.

\begin{figure}[ht!]
\begin{center}
\includegraphics[width=0.8\columnwidth]{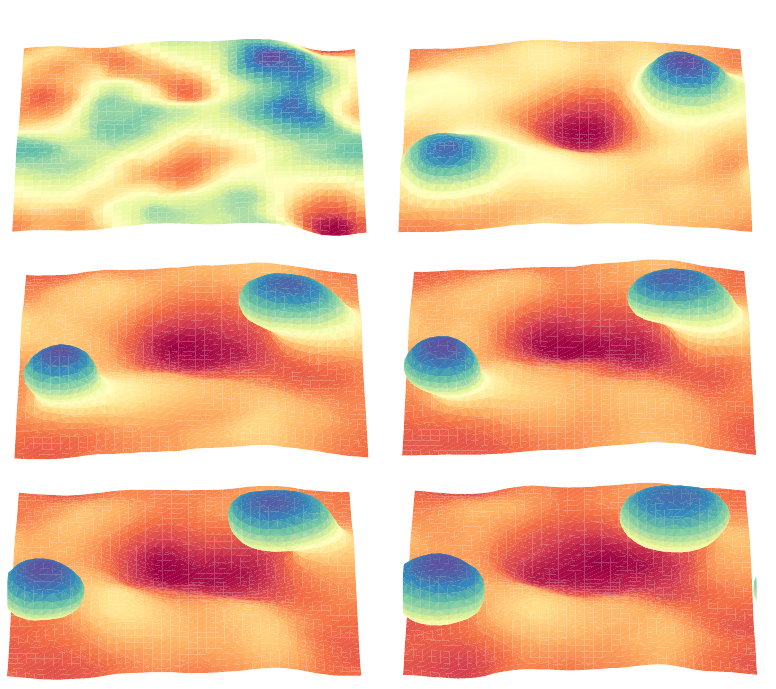}
\caption{  Snapshots of a simulation where the membrane generates vesicles. From a flat membrane the system generates two vesicles from the oscillations produced by the temperature. The parameters used $T=3.6\cdot10^{-5} (IU)$, $\epsilon=1$, $\kappa=1$, $\kappa_G=-2$ and $C_0=-0.25$. The color map of the membranes is present to aid height visualization. Red color means the membrane goes down and blue color means the membrane goes up.}
\label{FIG:vesiculation_1}
\end{center}
\end{figure}

With high enough temperature, we change into the vesiculation phase, where we can obtain vesicles from a flat membrane like in {Figure \ref{FIG:vesiculation_1}}. If we keep the simulation running, we can see vesicles keep forming until they fill up all the simulated space. Although the simulated vesicles after detachment do not conserve area nor volume perfectly the phenomenology of this system is all contained by the presented model for this kind of geometry.
It has been proven that even a membrane with a geometry unfavourable for fission, if it is combined with  temperature, one can obtain fission and vesicle formation.

{\bf Vesiculation phase diagram}

The only remaining question is how the phase space for vesicle formation and fluctuating  looks as a function of the free parameters. 
For that, we analyse the parameters that play an crucial role in whether fission happens or not. From the eq. (\ref{eq:kg_vs_k_T}) one sees that  the free parameters are: $T$, $\kappa_G/\kappa$, and $C_0$.
{Let us not forget that the real spontaneous curvature is slightly bigger than $C_0$ by $c_0 = \sqrt{2}C_0$. }

\begin{figure}[!ht]
\begin{center}
\includegraphics[width=.8\columnwidth]{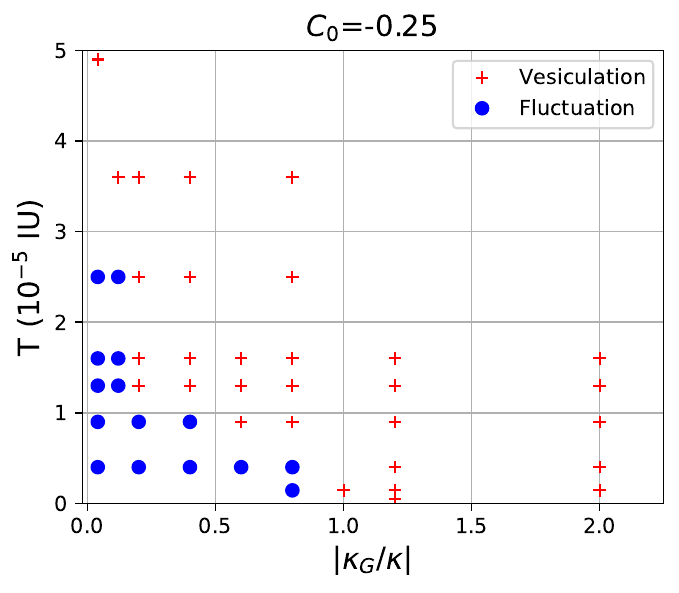}
\includegraphics[width=.8\columnwidth]{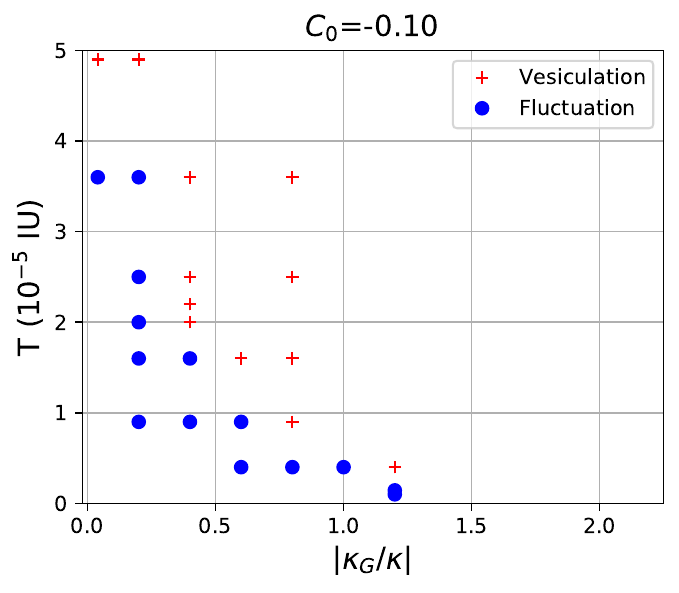}
\caption{ Phase diagram of vesicle formation from a flat membrane depending on temperature $T$ and the energy moduli ratio $\kappa_G/\kappa$ for a fixed spontaneous curvature $C_0$. The simulations had a size of $40\times40\times44$ and a membrane width of  $\epsilon=1$.}
\label{FIG:phase_diagram_c025}
\end{center}
\end{figure}

In Figure \ref{FIG:phase_diagram_c025}, we obtain a phase diagram for  fixed spontaneous curvatures $C_0=-0.25$ and  $C_0=-0.10$. For low temperatures and low ratio $|\kappa_G /\kappa|$, we conserve fluctuation regime. On the contrary, when any of these quantities increases, we obtain vesicle formation. The spontaneous curvature value helps the vesiculation process. The bigger $C_0$ the bigger the drop in energy between a flat surface (with curvature zero) and the a spherical vesicle of literally any curvature value, as it will reduce the bending energy contribution $ \kappa (C-C_0)^2.$ We can see when comparing the two plots in {Figure \ref{FIG:phase_diagram_c025}} that when decreasing the spontaneous curvature to $C_0=-0.10$, there are areas in  the phase diagram that change phase. It is be more pronounced for lower energy moduli ratios of $\kappa_G/\kappa$. However, one notes that the changes in $C_0$ generate {\color{black} qualitatively} equal diagrams. Therefore, one can fix spontaneous curvature $C_0$ and explore only a phase diagram for the temperature $T$ and the ratio $|\kappa_G /\kappa|$. {\color{black} Here we are only studying if the system vesiculates at very long times. Thus there will be no mixed-states.}

\begin{figure}[!ht]
\begin{center}
\includegraphics[width=\columnwidth]{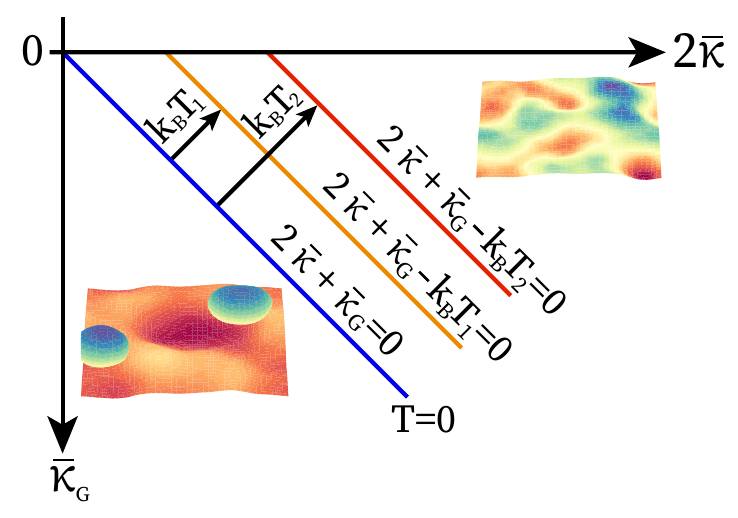} 
\caption{ Sketch of how the transition on the $\bar\kappa$ $\bar\kappa_G$ phase space moves with the thermal energy. 
Here we are using the original moduli $\bar\kappa$ and $\bar\kappa_G$ which have units of energy instead of the internal expressions used in the phase field model.
The blue line corresponds to the condition $- \bar\kappa_G >  2\bar\kappa$ which is the case of eq (\ref{eq:kg_vs_k_T}) for $T=0$. The other lines correspond to increasing temperature where $0<T_1<T_2$.  }
\label{FIG:diagrama_k_k}
\end{center}
\end{figure}

In a deep analysis, the results in Figure \ref{FIG:phase_diagram_c025} show that there is an energy barrier that one has to surpass to obtain vesicle formation. This energy barrier is being surpassed with the temperature. For any combination of $C_0$ and $|\kappa_G/\kappa|$ there is an energy barrier of a given magnitude, which decreases when increasing either $C_0$ or $|\kappa_G/\kappa|$. Thus, for a given energy barrier, as it is shown in the phase diagrams of the {Figure \ref{FIG:phase_diagram_c025}}, one sees how a minimum temperature is needed to be able to pass from the fluctuating phase into the vesiculating phase. Also, it is easy to see that from a given $\kappa_G$ on-wards vesiculation happens for extremely low temperatures.This happens as the value of $\kappa_G$ gets closer and surpasses the condition for spontaneous vesicle generation $- \bar  \kappa_G > 2\bar \kappa$. 
Thus, the region where one sees a transition between fluctuating and vesiculating  lies in the values  $- \bar \kappa_G < 2\bar \kappa$, when increasing $T$ and  where the Gaussian modulus is not big enough to fulfil spontaneous vesiculation. Otherwise,  we find that if $k_BT$ is lower than the energy barrier, we shall not find vesiculation. Therefore, a system with a given temperature, the phase transition between spontaneous vesiculation and stable topology is displaced. This displacement is proportional to $k_BT$ and can be represented by a displacement of the phase transition.
In Figure \ref{FIG:diagrama_k_k}, we show the phase diagram for $\bar \kappa_G$ and $\bar \kappa$ modulus, respectively. The distance to the transition like $- \bar \kappa_G > 2 \bar  \kappa$ can be interpreted as an energy barrier. The farther away a system is from the line the higher the barrier there is. Therefore, the higher the temperature is needed to make that point in the phase space transition to the vesiculation regime.\\

{\bf Estimating the real temperature}

For the estimation of the temperature, we will be using the Mean Square Displacement (MSD) $\langle | h|^2 \rangle$. To this end, we need to compute the exact position of the membrane over time. We do this by interpolating the position $\phi(x,y)=0$ to obtain $h(x,y)$. For a flat membrane is easy as we only need to go through the direction $z$ until finding the $\phi$ value closest to zero for each position $(x,y)$. This value $z$ will be our height $h(x,y)=z(x,y)$. With this information, we can use the following equation  \cite{Strey} 
\begin{equation}
\langle | h|^2 \rangle = \frac{k_B T}{2 \pi \bar \kappa} q_{min}^{-2} =   \frac{A \,k_B \, T}{8 \pi^3 \bar \kappa},
    \label{eq:MSDStrey}
\end{equation}
where $A=L^2$ is the surface area of the system {\color{black} we are simulating,} $L$ its length, and $q_{min}$ is the lower cutoff wave-vector and has a value $q_{min}=2\pi/\sqrt{A}$.

Using this expressions, we can compare the numerically obtained $\langle | h|^2 \rangle$ with the temperature $T$ and the bending modulus $\bar \kappa$. We can easily know $A$ and $\langle | h|^2 \rangle$ from a simulation and we can also fix $\bar \kappa$. Therefore, we can find the thermal energy of the system, from eq. (\ref{eq:MSDStrey}), is
\begin{equation}
    k_B \, T   =   8 \pi^3 \bar \kappa  \frac{ \langle | h|^2 \rangle }{A},
    \label{eq:KBTkappa}
\end{equation}
and it is independent on the scale that we are simulating. Here, the relevant value is the ratio between the surface area and the MSD. Moreover, the final temperature depends strongly on the rigidity $\bar\kappa$. The bending modulus can take a wide range of values, from $\bar \kappa \approx 10 k_B T_a$ to $\bar \kappa \approx 100 k_B T_a$ where $T_a=298$K. From this expression we can see that for a membrane with double the bending modulus $\bar \kappa$ we will need double the temperature to have the same MSD. 

In the simulations, we are using internal computational units (IU).
{ In the case of temperature these internal units are $ \bar \kappa/k_B $, the question is now which is the value in Kelvins. 
To do this we use} the mean square displacement of a number of simulations. The results of $\langle | h|^2 \rangle(\Delta x ^2)$  for the average of various simulations can be seen in Table \ref{tab:msd}. 
Here, we can see that the mean square displacement $\langle | h|^2 \rangle(\Delta x ^2)$ and the internal temperature $T$ are following eq. (\ref{eq:KBTkappa}) as the ratio of increase in $T$ is followed by the MSD. With these, we can compare the thermal energy with the MSD. Introducing the MSD for $T=0.9\cdot10^{-5}$IU in eq. (\ref{eq:KBTkappa}), one obtains $k_B \, T   \approx   0.012 \bar \kappa.$

\begin{table}[]
\centering
\begin{tabular}{|c|c|}
\hline
$T$ (IU)              & $\langle | h|^2 \rangle(\Delta x ^2)$ \\ \hline
$0.1 \cdot 10^{-5}$ & $0.008 $     \\ \hline
$0.4 \cdot 10^{-5}$ & $0.030 $     \\ \hline
$0.9 \cdot 10^{-5}$ & $0.075 $     \\ \hline
\end{tabular}
\caption{Mean square displacements $\langle | h|^2 \rangle$ obtained for different temperatures. The results of  $\langle | h|^2 \rangle$  are taken from the average of three different simulations at the same temperature. }
\label{tab:msd}
\end{table}

We note that the final simulated temperature depends on the bending modulus of what we will be simulating.
{ Using the ambient or room temperature $T_a = 293$K, the bending modulus } for soft vesicles made of a lipid like DOPC, is $\bar \kappa_{DOPC}=15 k_B T_a$ \cite{nagle2015true},  for vesicles or more rigid red blood cells, $\bar \kappa_{RBC}=70 k_B T_a$, and for diseased or drugged red blood cells it can go even higher to $\bar \kappa_{RBCd} = 120k_B T_a$   \cite{Paco}. Thus, the resulting thermal energy for each $\bar \kappa_{DOPC}$ and $\bar \kappa_{RBC}$ are $ k_B \, T \approx   0.18 k_B T_a$ and $ k_B \, T \approx 0.84 k_B T_a$, respectively. Therefore, depending on the bending modulus of the membrane $\bar\kappa$ at $T_{internal} = 0.9 \cdot 10^{-5}IU$, we are simulating around $T=54$K and $T=250$K. This is a wide range of temperatures available to simulate. The results presented here range mainly from $T_{internal} = 0.1 \cdot 10^{-5}IU$ up to $T_{internal} = 3.6 \cdot 10^{-5}IU$, which result in the wide temperature ranges represented in Table \ref{tab:temperatures}. Observe that our results do not seem to represent adequate temperature ranges for soft membranes, like the DOPC. Thus our simulations can be used to represent a range of real biological membranes of high bending rigidity $\bar \kappa$ with realistic temperatures. To study what happens to softer membranes, one would have to simply simulate higher temperatures that match the biological range of a given bending modulus $\bar \kappa$.

\begin{figure}[!ht]
\begin{center}
\includegraphics[width=0.8\columnwidth]{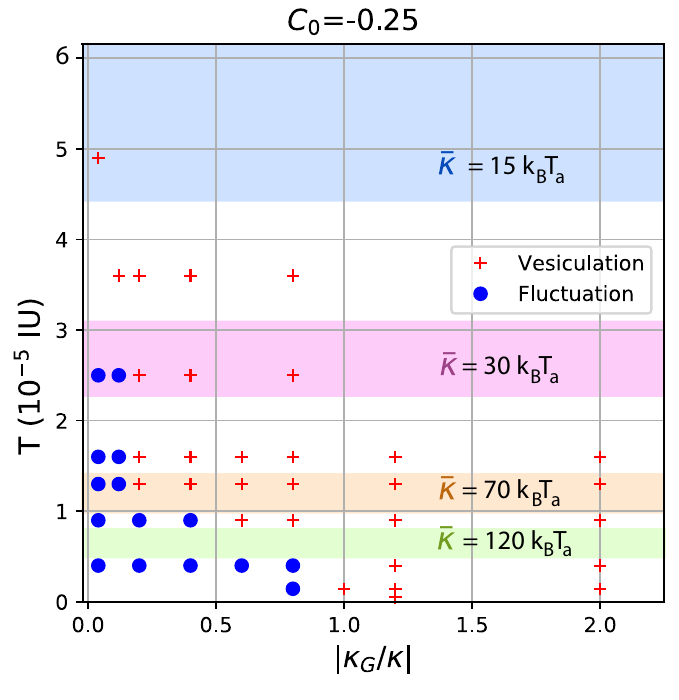} 
\caption{ Phase diagram of vesicle formation depending on temperature $T$ and the energy moduli ratio $\bar\kappa_G/\bar\kappa$ for a fixed spontaneous curvature $C_0$. In colours we find different regions where water is liquid depending on the bending modulus $\bar\kappa$. Simulations size of $40\times40\times44$ and membrane width  $\epsilon=1$.}
\label{FIG:phase_diagram_c025_T}
\end{center}
\end{figure}

Using Table \ref{tab:temperatures}, we can plot the region where water is in the liquid state in the previous phase diagram. In Figure \ref{FIG:phase_diagram_c025_T}, we show the dependency of the bending modulus in different regions of the phase diagram.
With this, we can look for a given bending if a transition to vesiculation or fluctuation regimes is possible by changing $T$, $C_0$, or $\bar \kappa_G$.

{ 
The length scales $\Delta x$  and time scale $\Delta t$ have to be compared to large membrane systems.
For example formation of GUVs, deformation on large lipid bilayers surfaces, and other large membranes which takes a physical scale of around 1-10 $\mu$m.
In these systems the time-scale of deformations will be in  the magnitude of 10s.
By comparison we can give a rough approximation of $\Delta x \sim 1 \mu$m and  $\Delta t \sim 10^{-2} $s. 
So here we are talking about systems of size around $40\times40\mu$m.
Looking at our computed $C_0$ the resulting vesicles in our simulations would be around $R_0 \sim 5 \mu$m for the results of Figure \ref{FIG:phase_diagram_c025_T}.

}

{ If we compare our results with DOPC systems, for example, which has a rather large spontaneous curvature of around $C_0 \approx 0.1/nm$ \cite{kollmitzer2013monolayer}.
In these cases the resulting vesicles generated by spontaneous vesiculation would have a size around $R_0 \sim 10$nm.
However, our results have to be compared to large systems as systems of nanometric length-scales require to take into account the monolayers themselves. 
For such scales one should use molecular dynamics or other microscopic approaches. 
}

\begin{table}[]
\centering
\begin{tabular}{c|c|c|c|}
\cline{1-4}
\multicolumn{1}{|c|} {T (IU)} & $0.9\cdot10^{-5}$  & $1.6\cdot10^{-5}$  & $3.6\cdot10^{-5}$  \\ \hline
\multicolumn{1}{|c|}{$\kappa=15 k_B T_a$} & 54 K                   & 96 K                  & 216 K                 \\ \hline
\multicolumn{1}{|c|}{$\kappa=30 k_B T_a$} & 108 K                  & 192 K                 & 432 K                 \\ \hline
\multicolumn{1}{|c|}{$\kappa=70 k_B T_a$} & 250 K                  & 444 K                 & 1000 K                 \\ \hline
\multicolumn{1}{|c|}{$\kappa=120 k_B T_a$} & 460 K                  & 850 K                 & 1850 K                 \\ \hline   
\end{tabular}
\caption{Temperatures from $IU$ to Kelvin for different internal values and different rigidities measured in $k_B T_a$ where $T_a=298$K and $k_B$ is the Boltzmann constant.}
\label{tab:temperatures}
\end{table}

{\textbf{Energy evolution}}

\begin{figure}[h!]
\begin{center}
\includegraphics[width=0.75\columnwidth]{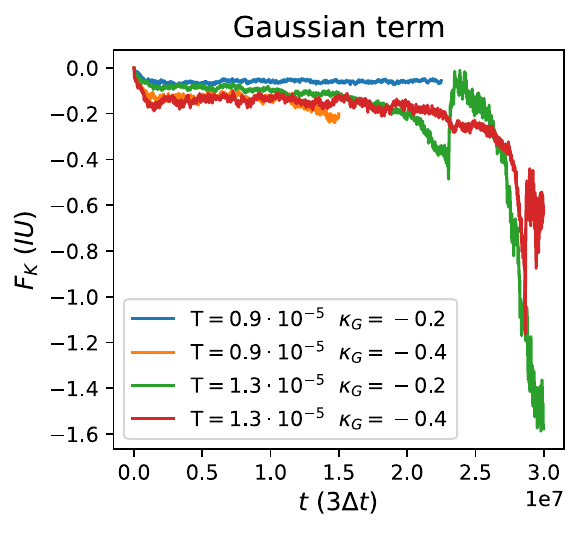} 
\includegraphics[width=0.75\columnwidth]{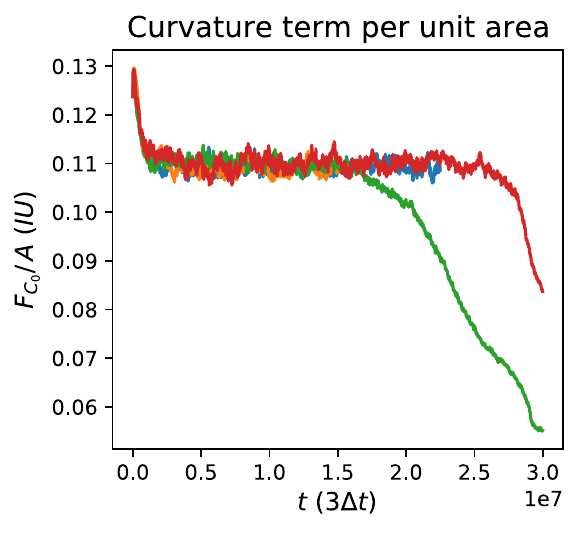} 
\caption{ Evolution over time of the two  energy contributions for a series of simulations, in some there is vesiculation and in some there is not. From the bending free energy we have the Gaussian energy term $F_K$ and the Curvature term energy density $F_{C_0}/A$. The curves that remained flat (blue and orange) remain on the fluctuating phase while the curves that decrease greatly (green and red) transition to the vesiculation phase.
{ The energy is measured in IU=$\bar \kappa$ and the energy per unit area in IU=$\bar \kappa / \Delta x ^2$  }
}
\label{FIG:energy_evol}
\end{center}
\end{figure}

We study the time evolution of both the Gaussian and the curvature energy contributions, represented in {Figure \ref{FIG:energy_evol}}. Here, one can see how for simulations where there is no vesiculation the energy curves remain flat over time, while for vesiculating systems the energy contributions decrease. However, looking at values that both terms take, we can see how the relative change in the Gaussian contribution is much greater than the one for the {\color{black} energy} Curvature term {\color{black} $F_{C_0}$ }. Therefore, when a system is at the vesiculating phase, the dynamics of the system are  driven mainly through the decrease in the Gaussian energy term $F_K$ (see eq. (\ref{eq:FG})). This energy term starts decreasing once the first vesicles fission from the main membrane, but as there is an energy barrier to surpass, the simulations that do not have a high enough temperature cannot get to lower their Gaussian energy term. 

In Figure \ref{FIG:energy_evol}, we obtain the curvature energy density $F_{C_0}/A$ by using eq. (\ref{eq:FC0}) and the surface area of the membrane.
This term in general decreases or remains constant, depending on whether we have vesicle formation or not. 
In simulations where we have a normal vesiculation process, the vesicles formed get a curvature closer to $C_0$ than when they where flat ($C=0$) and thus reduce the term $(C-C_0)^2$ from the free energy. 
Our interest to study the energy density of the curvature term instead of the total energy term is because as the surface area of the membrane increases over time the curvature energy contribution will also increase over time. 
This happens because in the end one has an energy cost in the curvature term of the free energy, Thus, any increase in surface area increases the curvature term of the free energy. 
Meanwhile, the Gaussian energy term, due to the Gauss-Bonnet theorem, is proportional to the number of objects and holes.
In conclusion the best way to understand the evolution of the membrane in a vesiculation scenario where area is not conserved is to plot the complete Gaussian term and the Curvature term per unit area.

\begin{figure}[h!]
\begin{center}
\includegraphics[width=0.8\columnwidth]{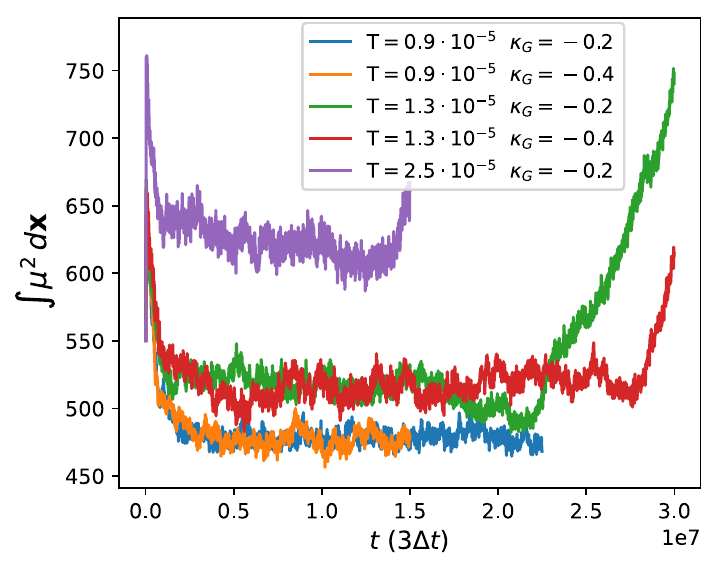} 
\caption{ Evolution over time of the chemical potential squared or $\mu^2$ for different simulations. In the plot there are both simulations where vesiculation occurs and some simulations where it does not.
Curves correspond to the same simulations as in {Figure \ref{FIG:energy_evol}} with the addition of the purple curve.
}
\label{FIG:mu_evol}
\end{center}
\end{figure}

Finally, we study the chemical potential term squared $\int \mu ^2 dV $, where the chemical potential is defined with the functional derivative $\mu = {\delta F_{C_0}}/{\delta \phi}$. This term shows the rate of change in the membrane. Looking at  $\mu^2$, in {Figure \ref{FIG:mu_evol}}, we observe that there is a flat plateau when the membrane is in the fluctuating state. The value of this plateau is clearly related to the temperature of the system, as all simulations at the same temperature overlap their plateaus. In cases where there will be vesiculation, this term starts climbing up as now there is an additional contribution to deformation other than temperature.

\begin{figure}[h!]
\begin{center}
\includegraphics[width=0.75\columnwidth]{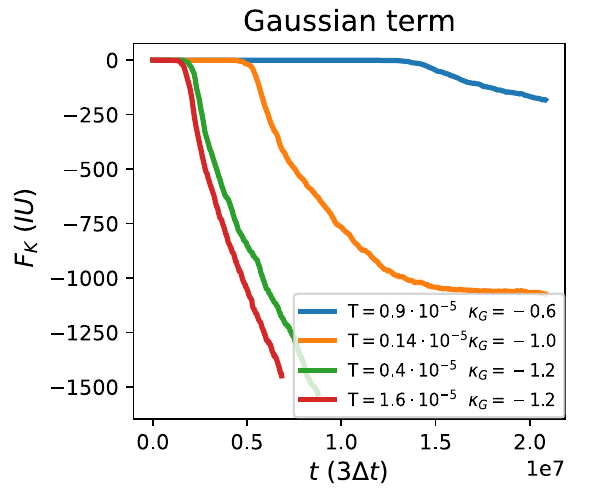} 
\includegraphics[width=0.75\columnwidth]{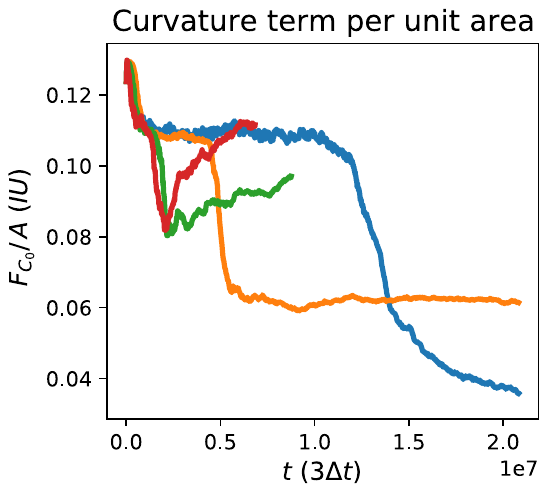} 
\caption{ Evolution over time of the two energy contributions for a series of simulations. From the bending free energy we have the Gaussian energy term and the Curvature term. In all simulations there is vesiculation and the main change in the rate at which the energy changes is mediated by the Gaussian modulus $\kappa_G$.
}
\label{FIG:energy_evol_kappas}
\end{center}
\end{figure}

Different values of $\kappa_G$ will make so that the energy contribution of each vesicle created is bigger with increasing $|\kappa_G|$. This changes the rate of change over time for the Gaussian energy contribution. Thus the slope of the Gaussian energy term over time will depend on both temperature $T$ and the Gaussian modulus $\kappa_G$. However,  in {Figure \ref{FIG:energy_evol_kappas}}, one can see how by doubling  the value of $\kappa_G$ from the blue curve to the green one the slope of the Gaussian energy curve increases drastically, even with a lower $T$. Thus in this figure, one can see that the rate of change over time of the energy is very dependent on the Gaussian modulus $\kappa_G$. The temperature can be seen to change also that rate between the two simulations of {Figure \ref{FIG:energy_evol_kappas}} that share the same modulus $\kappa_G=-1.2$. However, $\kappa_G$ is more influential on the energy terms than the temperature.

For the curvature energy density in {Figure \ref{FIG:energy_evol_kappas}}, we can see a similar behaviour as in previous plots for most values of $\kappa_G$. 
However, in some cases the curvature term climbs back again. In these cases the $\kappa_G$ is so big that its dominating over the curvature term and the vesicles deform in non-spherical ways to try to fit more vesicles at the cost of increasing the curvature energy term. Also, one can see how after a long time of vesiculation the Gaussian and Curvature energy terms stop changing and flattens again. This happens because in these simulations that have been left to run a long time the system reaches a point where all the space is full of vesicles and there is no room to produce any more.\\

{\bf Positive Gaussian modulus }

\begin{figure}[ht!]
\begin{center}
\includegraphics[width=.48\columnwidth]{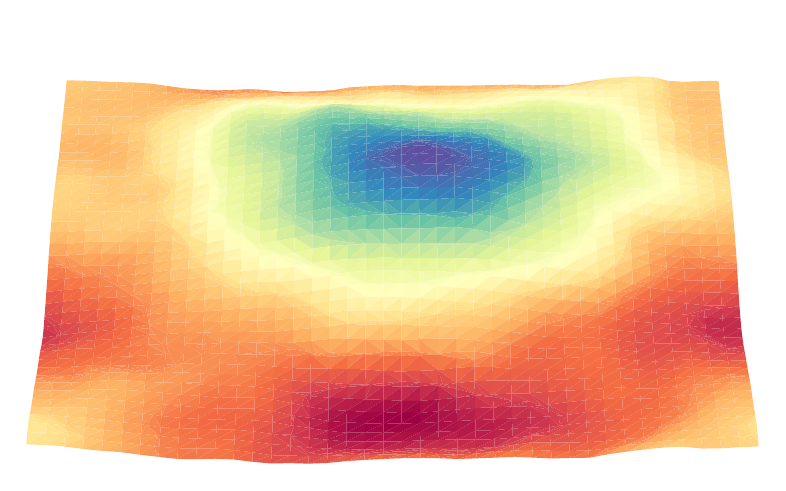}
\includegraphics[width=.48\columnwidth]{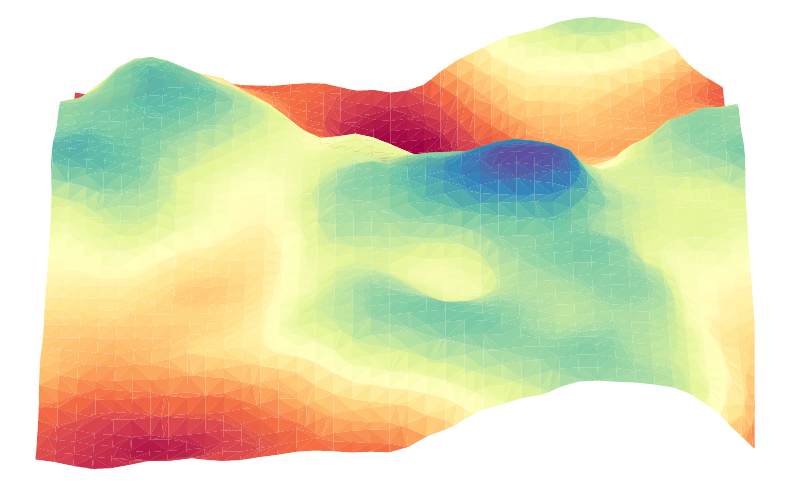}
\caption{ \textbf{Left:} Membrane fluctuating with T=$6.4 \cdot 10^{-5}$IU, $\kappa_G =0.4$, and $C_0=-0.25$. \textbf{Right:} Membrane fluctuating with T=$256 \cdot 10^{-5}$IU, $\kappa_G =0.4$, and $C_0=-0.25$.
Little matters the intensity of the noise, for positive Gaussian modulus we find no topological transitions.
For very high noises the membrane curves and stays curved, but does not go further than that.
}
\label{FIG:positiveK}
\end{center}
\end{figure}

Changing the sign of Gaussian  modulus $\kappa_G$ to a positive, one completely changes the results.Getting the system to transition from the fluctuating regime seems impossible, even for unrealistically high noise, the Gaussian contribution can hold the membrane together. The morphology of the membrane also changes, for very big noises like in {Figure \ref{FIG:positiveK}}. We observe huge deformations and valleys much bigger than the size of the vesicles produced with negative $\kappa_G$. As high as these deformations are and as high as the temperature, the simulation never shows fission or handles.

Some of the temperatures introduced in Figure \ref{FIG:positiveK} are two orders of magnitude over the previously simulated temperatures. Nonetheless, it seems to not matter how much thermal energy we introduce the topology of the simulation never changes.  The Gaussian term seems to be holding the membrane together, as for simulations with $\kappa_G=0$ at these extremely high temperatures the system breaks down. Usually the Gaussian contribution is thought of only influencing the system when small membrane necks and other shapes close to topological transitions happen. However, the results obtained seem to suggest that the Gaussian term is changing radically how the membrane deforms at all stages and membrane geometries, at least for thermal fluctuations.

The lack of vesiculation makes sense given the Gauss-Bonnet theorem, when $\kappa_G$ is taken positive, predicts that increasing the number of holes is what would lower the energy of the system. 
This however, does not happen. 
Even if the configuration is hard to achieve, the high temperatures used should be giving the necessary energy to produce any membrane deformation necessary. 
Therefore, maybe the most intriguing of all is not reaching a self-connected membrane like we see in a previous work \cite{rueda2021gaussian}. 
One would expect with this level of noise to go through topological transitions but in the direction of increasing the number of holes and that decreasing the Euler characteristic and the Gaussian contribution. 
Thus, it seems that a flat membrane geometry is more difficult to generate holes and passages than to produce vesicles.\\


{\bf Turning off the temperature}

The temperature is necessary to start the fission process. However, there is no need to maintain it after a certain time.
Even for simulations where there has still not been a single fission event, if the membrane is already curved enough due to the temperature, even after turning off the noise it will end in multiple fission events.

\begin{figure}[h!]
\begin{center}
\includegraphics[width=.5\columnwidth]{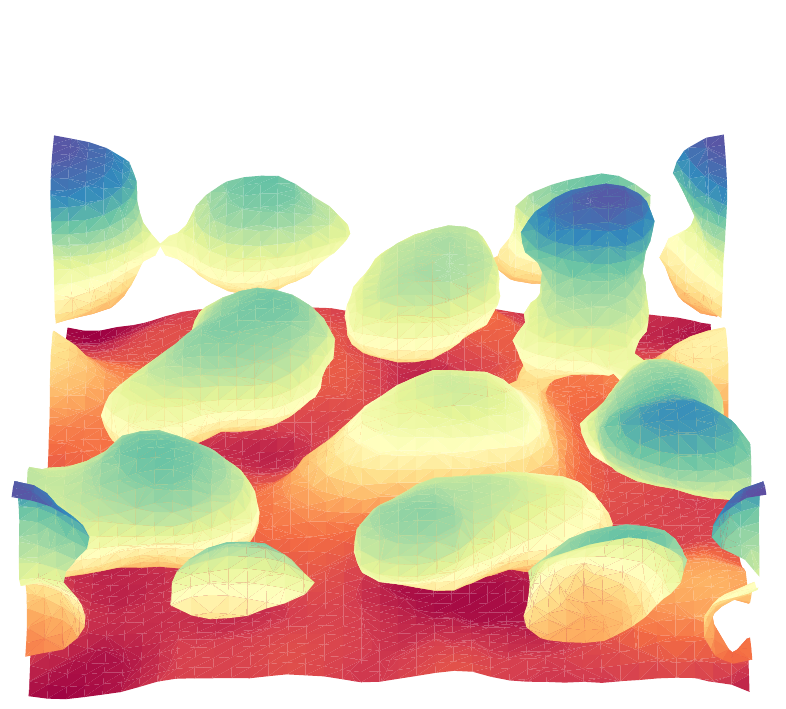}
\includegraphics[width=.47\columnwidth]{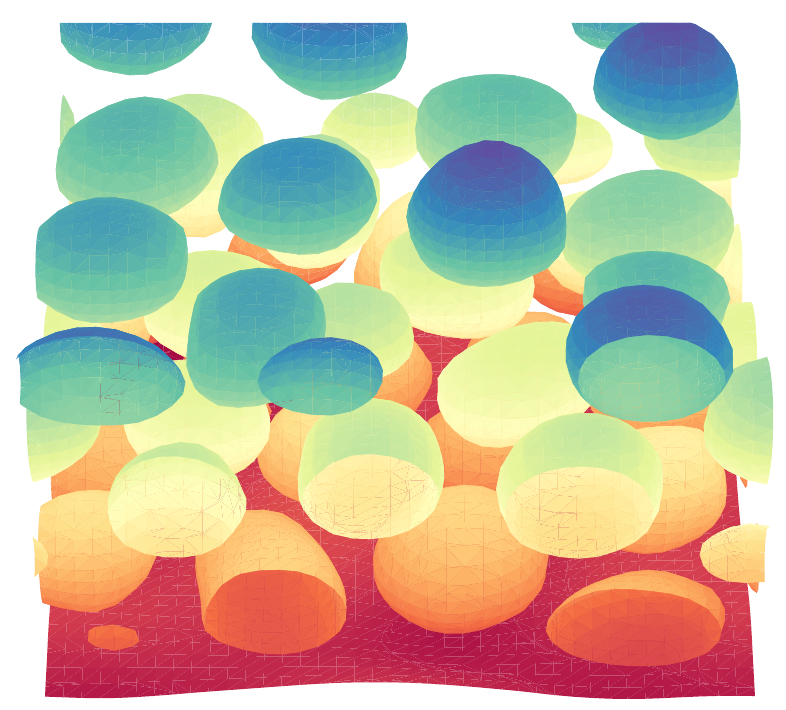}
\caption{\textbf{ Left: } At very high $T$ the membrane shape has more rough edges and might stick to new vesicles if there is no more room left. \textbf{Right:} Even after turning the noise off the initially flat membrane keeps generating vesicles, and the vesicles take their preferred spherical shape.
This simulation was at $T=144\cdot 10^{-5}$ IU, $\kappa_G =-0.4$, and $C_0=-0.25$. }
\label{FIG:intermitent}
\end{center}
\end{figure}

Another consequence of the noise is that the vesicle shape and dynamics are slightly rough.
One can see edges that one would not expect under normal circumstances  as the resulting vesicles sometimes deform into rough shapes.
This can be seen in the left plot of {Figure \ref{FIG:intermitent}}.
This happens because the vesicles are rather small and when in  comparison with the thermal fluctuations the latter are big enough to deform the vesicles a lot.
As the fluctuations are rather big at the size scale of a vesicle, it also makes more difficult to conserve area and volume and, most importantly, to be dominated  the bending.
However as stated earlier, the point of these simulations is not to have a rigorous description of the vesicles after the fission but to study the process that leads to fission.

To prove that this weird behaviour after fission is due to the noise and not to problems in the computation or the model, we study what happens if we turn the noise off mid-way through a simulation.
One can see that after an initial kickoff where the membrane starts to be deformed enough to generate vesicles, even after turning off the temperature one will still get vesiculation.
We can see this case in {Figure \ref{FIG:intermitent}}, where after reaching a point where the membrane starts to generate vesicles we turn off the temperature and the vesicles behave more properly and vesiculation continues.
This happens because the fluctuations made the system reach a much better geometry to fission and until reaching the steady state more vesicles will be formed only due to the Gaussian energy contribution.

{\color{black} All this reflects the fact that temperature suffices to provide the system with enough energy to vesiculate. }

\section{Discussion} 

The conditions for spontaneous fission are present, a flat membrane will fission only in the presence of temperature, as one needs a trigger starting the process. Moreover, we see that through temperature, membranes that do not comply the original condition $- \bar \kappa_G > 2\bar \kappa$ can exhibit spontaneous fission. This is of great importance as membranes never are in a $T=0$ scenario. 
In the end the thermal energy related to this noise by $E_T=k_B T$ is helping the system overcome an energy barrier $\Delta E$ to achieve vesiculation.

There are two possible states in the presence of temperature for the membrane: either its only fluctuating or its undergoing vesiculation. 
The phase diagrams show clearly how the spontaneous curvature $C_0$, temperature $T$, and both Gaussian and bending moduli, $\kappa_G$ and $\kappa$, influence in the transition.  
While $C_0$ changes the position of the transition in the diagram, it does not change the overall shape of the transition region. 
This transition is clearly seen in the region where the condition for spontaneous vesiculation $- \bar \kappa_G > 2\bar \kappa$ is not met and it happens because the region where the condition is met requires minimal temperature to trigger the vesiculation and the closer one is to meeting this condition the lower $T$ is required for fission to start. 
The phase diagram  $(\kappa_G, \kappa)$ represented this displacement by changing the transition line as $ \bar \kappa_G = - 2 \bar  \kappa +k_BT$ (see Figure \ref{FIG:diagrama_k_k}). The results clearly show how higher spontaneous curvatures help the vesiculation process by lowering the energy barrier $\Delta E$ to overcome.

For membranes with a positive Gaussian modulus $\kappa_G$, there is no temperature high enough so that it can transition from the fluctuating regime. We do not observe the membrane use the thermal fluctuations to create neither vesicles nor handles or passages as we see in \cite{rueda2021gaussian}. One would expect with this level of noise to maybe go through topological transitions by increasing the number of holes. This points either to the flat geometry being bad to generate handles and passages or to the solution being unstable.

All the processes are driven mostly through the Gaussian energy term of the Helfrich free energy. This is seen in the relative change in the Gaussian energy term being greater than the energy term related to curvature as seen in Figure \ref{FIG:energy_evol}. The evolution of the Gaussian energy term seems to scale exponentially with the Gaussian rigidity $\kappa_G$ while the influence of temperature $T$ is much smaller (Figure \ref{FIG:energy_evol_kappas}). Also from a numerical point of view, there is value on working with the integral of the chemical potential squared $\int \mu ^2dx$, which clearly can be used to measure the temperature of the system.

These results can be exploited to measure the value of the Gaussian rigidity $\bar\kappa_G$ on real membranes. By introducing different proteins or substances that induce an spontaneous curvature $C_0$ and then exploring the whole available range of temperatures one should be able to find the transition from fluctuating to vesiculation. In the case that no transition is found, one could introduce a higher {\color{black} absolute value of} $C_0$ and explore again all the available range of temperatures. The temperature at which the transition will end up happening will depend on the value of $\kappa_G/\kappa$. Given that the value of the bending modulus (or rigidity) $\kappa$ is much easier to obtain experimentally than $\kappa_G$ this could be a method for obtaining the Gaussian energy modulus.

An open question that remains is whether the different molecular machinery involved in some fission processes (e.g. the dynamic superfamily \cite{Schmid,Hinshaw,Sweitzer} or the ESCRT machinery\cite{Bleck,Engelenburg}) work by lowering the observed energy barrier $\Delta E$ or simply produce fission regardless of $\Delta E$. There is the possibility of this depending on the method that each molecular motor involved in fission uses. All these processes have a chemical rate and this rate is also dependent on temperature, so experiments on determining the nature on whether they are dependent on the energetic of the system or not could be challenging.

\section*{Acknowledgements}

We acknowledge insightful discussions with F. Campelo, F. Monroy, J. Ignés-Mullol, I. Pagonabarraga, and R. Reigada.
A.F.G. acknowledges financial support from MINECO (Spain) project FIS2016-78883-C2-1-P. Likewise, J.R.R.A. thanks the support from DAGAPA-UNAM grant IA-100823. A.H.M. acknowledges financial support from Ministerio de Ciencia e Innovación (MICINN, Spain) project PID2019-1060636B-100. J.R.R.A. and
R.A.B. are grateful for the hospitality at the Universitat de Barcelona. R.A.B. was financially supported by Conacyt through project 283279.


\bibliography{references} 
\bibliographystyle{rsc} 
\end{document}


\begin{suppinfo}

\end{suppinfo}

\section*{Appendix}
{\bf Obtaining the phase field model}

To write the Canham-Helfrich free energy equation in terms of a phase field order parameter $\phi$ we will have to first define the total and Gaussian curvatures.
Having the surface normal vector $\hat n$ one can define then  the curvature tensor 
\begin{equation}
    Q_{ij} = \nabla_i  \, n_j,
    \label{eq:curvaturetensor}
\end{equation}
as  the normal vector $n$ changes along the surface are directly related to the curvature of a surface.

Now, using the curvature tensor $Q$ we can write the curvature $C$ and the Gaussian curvature $K$. The determinant of the curvature tensor $Q$ is always zero, but using its trace we can write the curvature \cite{campeloTh}
\begin{equation}
C= \text{tr}[Q].
    \label{eq:C}
\end{equation}
The Gaussian curvature  is a bit more complex being 
\begin{equation}
K  =  \sum_{i,j} \Big [ \Big(  Q_{ii}Q_{jj}-Q_{ij}^2 \Big) \frac{1-\delta_{ij}}{2}  \Big].
    \label{eq:K}
\end{equation}
The expression is focused on the non-diagonal elements of the tensor because it is related to the overall shape that the surface takes around a point and whether is plane, curved, or presents a saddle-splay shape.

To compute the simulations with a simple model that avoid the need of tracking the interface position we will use a phase field model.  Therefore, we define an order parameter $\phi(\textbf{x})$ that represents if the volume of fluid at the point $\textbf{x}$ corresponds to either external or internal fluid to the vesicle. In this article, we choose the values $\phi(\textbf{x})=+1$ for the internal fluid of the vesicle and $\phi(\textbf{x})=-1$ for the external fluid. With this order parameter,  we can write the curvature parameters as \cite{campelo06}  
\begin{equation}
C [\phi] =  \frac{\sqrt{2}}{\epsilon(1-\phi^2)}\Big( -\phi + \phi^3 -\epsilon^2 \nabla^2\phi \Big),
    \label{eq:C_PF}
\end{equation}
and, to compute $K$, we will be using the curvature tensor written for a phase field
\begin{equation}
Q_{ij} =  \frac{\sqrt{2}\epsilon}{1-\phi^2} \Big[ \partial_i\partial_j \phi + \frac{2\phi}{1-\phi^2}\partial_i \phi \partial_j \phi \Big],
    \label{eq:Q_PF}
\end{equation}
where $\partial_i$ refers to a derivative and $i,j$ can be any of our cartesian coordinates $(x,y,z)$. We will be computing the Canham-Helfrich free energy as a phase field which we will split into two different contributions $F={F_{C_0}}+{F_{G}}$, which consists  of the Curvature energy term and the Gaussian curvature term.

The minimisation of the Canham-Helfrich free energy  is computed numerically in this scheme. In this methodology the main goal is to change from a surface energy to a volumetric energy by defining an order parameter. For this we use a surface differential  expressed like
\begin{equation}
    dS = \frac{3}{4\sqrt{2}\epsilon}\Big ( 1-\phi^2 \Big)^2 dV.
    \label{eq:dS}
\end{equation}


{\bf Obtain ${\bf T_1}$ and ${\bf T_2}$}

The Gaussian energy term written as a phase field model involve many derivatives of the order parameter $\phi$. For simplicity, we have called them $T_1$ and $T_2$, here we have written explicitly as introduced in \cite{rueda2021gaussian}. The various derivatives of $\phi$ are written for simplicity as $\phi_i$ where $i$ is either $x$ $y$ $z$ 
\begin{equation}
   T_1 = \phi_{xx} \phi_{yy} + \phi_{xx}\phi_{zz} + \phi_{yy}\phi_{zz} - (\phi_{xy})^2 - (\phi_{xz})^2- (\phi_{yz})^2 ,
\end{equation}
and 
\begin{align}
   T_2 = \phi_{xx} (\phi_{y})^2 + & \phi_{xx}(\phi_{z})^2 + \phi_{yy} (\phi_{x})^2  + \phi_{yy} (\phi_{z})^2 + \phi_{zz} (\phi_{x})^2  + \phi_{zz} (\phi_{y})^2  \nonumber \\
  & - 2 \phi_x\phi_y\phi_{xy} - 2 \phi_x\phi_z\phi_{xz} - 2 \phi_y\phi_z\phi_{yz} .
\end{align}

\bibliography{references}